\documentclass[twocolumn,showpacs,preprintnumbers,amsmath,amssymb]{revtex4}
\usepackage{mathrsfs}

\usepackage{graphicx}
\usepackage{dcolumn}
\usepackage{bm}
\usepackage{amsmath}

\begin{document}


\title{Generation of two-mode entanglement between separated cavities}

\author{Pengbo Li}
\affiliation{State Key Laboratory for Mesoscopic Physics, Peking
University, Beijing 100871, China}

\date{March 16, 2008}

\begin{abstract}
We propose a scheme for the generation of two-mode entangled states
between two spatially separated cavities. It utilizes a two-level
atom sequentially  coupling to two high-Q cavities with a strong
classical driving field. It is shown that by suitably choosing the
intensities and detunings of the fields and coherent control of the
dynamics, several different entangled states such as entangled
coherent states and Bell states can be produced between the modes of
the two cavities.
\end{abstract}

\pacs{03.67.Mn, 42.50.Dv, 42.50.Pq} \maketitle

The preparation of quantum entangled states continues attracting
intense theoretical and experimental activities. These nonclassical
states not only are utilized to test fundamental quantum mechanical
principles such as Bell's inequalities \cite{Bell} but also plays a
central role in practical applications of quantum information
processing \cite{quantum_information} such as quantum
computation\cite{Shor,Prl-79-325}, quantum
teleportation\cite{Prl-70-1895}, and quantum
cryptography\cite{Prl-67-661}. In quantum optics the generation of
various nonclassical states especially entangled states of
electromagnetic fields is a central topic \cite{quantum_optics}. In
the context of cavity QED \cite{Kimble,RMP-73-565,Sci298},
experimental realizations of the entanglement between two different
modes sharing a single photon or two polarized photons in a cavity
have been reported \cite{pra-64-050301,Sci317}. In recent years,
great effort has been put into preparation of the Schr\"{o}dinger
cat states \cite{Prl-77,Sci272}, where the extreme cat states are
reduced to mesoscopic quantum states with classical counterparts,
i.e., coherent states. Recently, a method of generating entangled
coherent states \cite{pra-40} between different modes in a cavity
has been presented \cite{Prl-90-027903}. There are also several
other proposals for producing entangled field states between
different cavities
\cite{pra-67-012325,pra-71-053814,pra-74-031801,pra-74-044301,eprint}.

In this paper, we propose a scheme for generating two-mode entangled
states such as entangled coherent states and Bell states between two
distant cavities. A two-level atom is sent sequentially into two
spatially separated cavities, assisted by a strong classical driving
field \cite{Prl-90-027903}. We demonstrate that by suitably choosing
the intensities and detunings of the fields and coherent control of
the dynamics, several different entangled states such as entangled
coherent states and Bell states can be produced between the two
cavity modes. These entangled states of fields can have both
fundamental applications in quantum mechanics and practical
applications in quantum information processing. With presently
available experimental setups in cavity QED, this protocol can be
implemented.

This proposal consists of two distant high-Q cavities and a
two-level atom, as sketched in Fig. 1.
\begin{figure}[h]
\centering
\includegraphics[bb=129 537 345 788,totalheight=2.5in,clip]{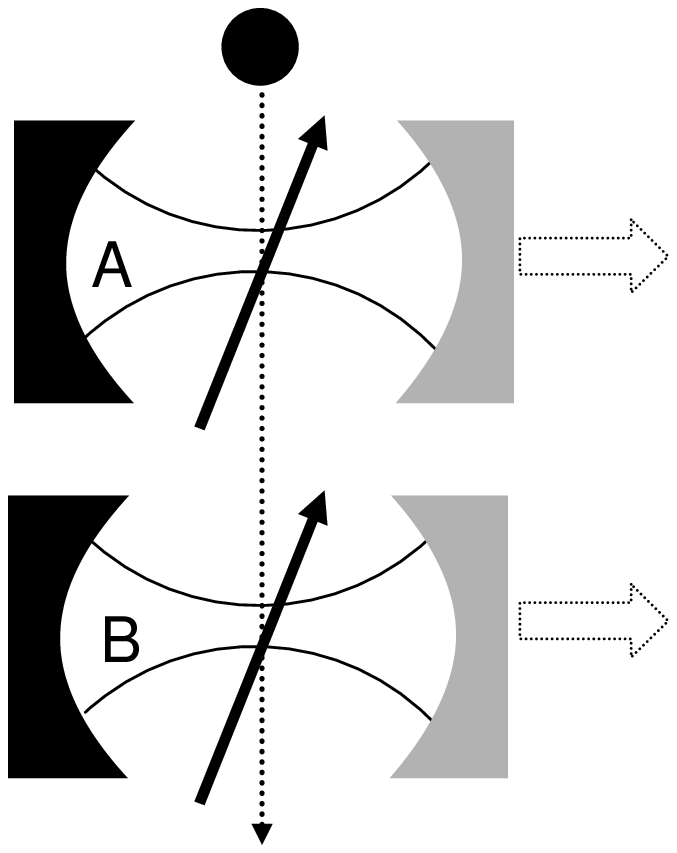}
\caption{Proposed experimental setup. A two-level atom sequentially
couples to two distant cavities A and B, driving by a strong
classical field. }
\end{figure}
The atom sequentially couples to the cavities A and B (with cavity
modes of frequencies of $\nu_A$ and $\nu_B$), driving by a strong
classical field (frequency $\omega _{L}$). The ground state of the
atom is labeled as $\vert g\rangle$, and the excited state as $\vert
e\rangle$. In each cavity, the transition $\vert
g\rangle\leftrightarrow \vert e\rangle$ (transition frequency
$\omega_0$) is coupled by the cavity mode with the coupling
constants $g_A$ and $g_B$, respectively. Furthermore, a strong
classical field drives the same transition with a Rabi frequency
$\Omega_A (\Omega_B)$ in each step. The associated Hamiltonian for
the dynamics in each cavity under the dipole and rotating wave
approximation is given by (let $\hbar=1$)
\begin{eqnarray}
\label{H}
H_j=&& \omega_0 \sigma^\dag \sigma+\nu_j \hat{a}^\dag_j \hat{a}_j\nonumber\\
&&+\Omega_j(e^{-i\omega_Lt}\sigma^\dag
+e^{i\omega_Lt}\sigma)\nonumber\\
&&+g_j(\sigma^\dag \hat{a}_j+\sigma\hat{a}^\dag_j), (j=A,B).
\end{eqnarray}
Where $\sigma=\vert g\rangle\langle e\vert$, and $\sigma^\dag=\vert
e\rangle\langle g\vert$ are the atomic transition operators;
$\hat{a}_j$ and $\hat{a}^\dag_j$ are the annihilation and creation
operators with respect to cavity $j$. In the following we assume
that $\Omega_A$, $\Omega_B$, $g_A$, and $g_B$ are real for
simplicity.

The Hamiltonian of Eq. (\ref{H}) can be changed to a reference frame
rotating with the driving field frequency $\omega _{L}$,
\begin{eqnarray}
H_j^L=&\Delta \sigma^\dag \sigma+\delta_j\hat{a}^\dag_j
\hat{a}_j+\Omega_j(\sigma^\dag +\sigma)\nonumber\\
&+g_j(\sigma^\dag \hat{a}_j+\sigma\hat{a}^\dag_j), (j=A,B),
\end{eqnarray}
with $\Delta=\omega_0-\omega_L$ and $\delta_j=\nu_j-\omega_L$. In
the following we will set $\Delta=0$ for simplicity. We now switch
to a new atomic basis $\vert \pm\rangle=(\vert g\rangle\pm\vert
e\rangle)/\sqrt{2}$. In the interaction picture with respect to
$H^j_{0}=\delta _{j}\hat{a}^{\dag }\hat{a}_j+\Omega_j
(\sigma^\dag+\sigma)$, we have the following Hamiltonian
\begin{eqnarray}
\label{H_I} H_j^I=&&\frac{g_j}{2}(\vert +\rangle\langle +\vert-\vert
-\rangle\langle -\vert+e^{2i\Omega_j t}\vert +\rangle\langle
-\vert\nonumber\\
&&-e^{-2i\Omega_j t}\vert -\rangle\langle
+\vert)\hat{a}_je^{-i\delta_jt}+\mbox{H.c.}
\end{eqnarray}
The Hamiltonian (\ref{H_I}) is the starting point in the following
discussions, from which we show that a variety of entangled states
of the two distant cavities can be generated through suitably
choosing the detunings and light intensities.

We first show how to produce the entangled coherent states between
the two cavities. Assume that $\Omega_A (\Omega_B)\gg \{g_A,\delta_A
(g_B,\delta_B)\}$, this strong driving condition can allow one to
realize a rotating-wave approximation and neglect the fast
oscillating terms. Now $H_j^I$ reduces to
\begin{eqnarray}
\label{H4} H_j^I=&&\frac{g_j}{2}(\vert +\rangle\langle +\vert-\vert
-\rangle\langle
-\vert)(\hat{a}_je^{-i\delta_jt}+\hat{a}_j^\dag e^{i\delta_jt})\nonumber\\
=&&\frac{g_j}{2}(\sigma^\dag+\sigma)(\hat{a}_je^{-i\delta_jt}+\hat{a}_j^\dag
e^{i\delta_jt}).
\end{eqnarray}
If we choose $\delta_j=0$, this Hamiltonian corresponds to the
simultaneous realization of Jaynes-Cummings (JC) \cite{JC}and
anti-Jaynes-Cummings (AJC) interaction in each cavity. The evolution
operator for the system is given by
\begin{eqnarray}
\label{U1}
U_j(t)&=&e^{-iH_j^It}\nonumber\\
 &=&e^{-\frac{ig_jt}{2}(\sigma^\dag+\sigma)(\hat{a}_j+\hat{a}_j^\dag
)}\nonumber\\
&=&\hat{D}(\alpha_j)\vert
+\rangle\langle+\vert+\hat{D}(-\alpha_j)\vert -\rangle\langle-\vert,
\end{eqnarray}
with
$\hat{D}(\alpha_j)=e^{\alpha_j\hat{a}_j^\dag-\alpha_j*\hat{a}_j}$,
and $\alpha_j=-ig_jt/2$. Assume that at the time $t=0$ the system is
prepared in the ground state $\vert g\rangle \vert 0\rangle_A\vert
0\rangle_B=(\vert +\rangle+\vert -\rangle)\vert 0\rangle_A\vert
0\rangle_B/\sqrt{2}$, i.e., the atom stays in $\vert g\rangle$, and
the two cavities are in the vacuum states. The atom enters cavity A
and undergoes the dynamics of Eq. (\ref{U1}). The evolved state
after a time $t_A$ will be
\begin{equation}
\frac{1}{\sqrt{2}}(\vert +\rangle\vert \alpha\rangle_A+\vert
-\rangle\vert -\alpha\rangle_A)\vert 0\rangle_B,
\end{equation}
where $\alpha=-ig_At_A/2$ for the case of $\delta_A=0$. This
microscopic-mesoscopic entangled state is the so-called
Schr\"{o}dinger cat state. After an interaction time $t_A$ in cavity
A, the atom enters cavity B. It will also undergoes the dynamics of
Eq. (\ref{U1}). After a time $t_B$, the system consisting of the two
cavities and the atom will evolve into
\begin{equation}
\label{entanglement1} \frac{1}{\sqrt{2}}(\vert +\rangle\vert
\alpha\rangle_A\vert \beta\rangle_B+\vert -\rangle\vert
-\alpha\rangle_A\vert -\beta\rangle_B),
\end{equation}
with $\beta=-ig_Bt_B/2$ for the case of $\delta_B=0$. Equation
(\ref{entanglement1}) describes a tripartite entangled state
involving one microscopic and two mesoscopic systems. If we measure
the atomic state in the bare basis $\{\vert g\rangle,\vert
e\rangle\}$, we can obtain the entangled coherent states of the
fields in the interaction picture
\begin{equation}
\mathscr{N}_{AB}^{\pm}(\vert \alpha\rangle_A\vert
\beta\rangle_B\pm\vert -\alpha\rangle_A\vert -\beta\rangle_B),
\end{equation}
where $\mathscr{N}_{AB}^{\pm}$ is the normalized factor. It has been
shown that these states can be utilized as an important tool in the
field of quantum information. Here we propose to generate the
entangled coherent states between two distant resonators.

We next show that using the interaction described by Eq. (\ref{H_I})
one can generate the maximally entangled state $1/\sqrt{2}(\vert
0\rangle_A\vert 1\rangle_B+\vert 1\rangle_A\vert 0\rangle_B)$
between the two cavities. If we choose $\delta_j=2\Omega_j$ and
$|\delta_j|\gg g_j$, then we can bring the Hamiltonian (\ref{H_I})
to the JC interaction in the $\vert \pm\rangle$ atomic dressed basis
\begin{equation}
\label{JC} H_j^{JC}=\frac{g_j}{2}(\vert
+\rangle\langle-\vert\hat{a}_j+\vert
-\rangle\langle+\vert\hat{a_j}^\dag).
\end{equation}
Assume at $t=0$, the system stays in $\vert +\rangle\vert
0\rangle_A\vert 0\rangle_B$. Then after an interaction time
$t_A=\pi/(2g_A)$ in cavity A, the system will evolve into
\begin{eqnarray}
\label{entanglement2} \frac{1}{\sqrt{2}}(\vert +\rangle\vert
0\rangle_A-i\vert -\rangle\vert 1\rangle_A)\vert 0\rangle_B.
\end{eqnarray}
Subsequently, the atom enters cavity B, and undergoes the dynamics
of Eq. (\ref{JC}). Then the atom-field state will be
\begin{widetext}
\begin{eqnarray}
\frac{1}{\sqrt{2}}(\cos(g_Bt/2)\vert +\rangle\vert
0\rangle_B-i\sin(g_Bt/2)\vert -\rangle\vert 1\rangle_B)\vert
0\rangle_A-\frac{i}{\sqrt{2}}\vert-\rangle\vert1\rangle_A\vert0\rangle_B.
\end{eqnarray}
\end{widetext}
If the interaction time $t_B=\pi/g_B$ is taken, the final state will
be
\begin{eqnarray}
\frac{1}{\sqrt{2}}(\vert 0\rangle_A\vert 1\rangle_B+\vert
1\rangle_A\vert 0\rangle_B)\vert -\rangle,
\end{eqnarray}
where a common phase factor $-i$ has been discarded. Clearly, at
this time the atomic state has been factorized out and the modes of
two distant cavities end up in the EPR pair state\cite{epr}
\begin{equation}
\label{Bell1} \frac{1}{\sqrt{2}}(\vert 0\rangle_A\vert
1\rangle_B+\vert 1\rangle_A\vert 0\rangle_B).
\end{equation}
From the viewpoint of quantum information, this is a maximally
entangled state of two qubits stored in the modes of two distant
cavities.

Now we consider the case of generating the entangled state
$1/\sqrt{2}(\vert 0\rangle_A\vert 0\rangle_B-\vert 1\rangle_A\vert
1\rangle_B)$ between the two cavities. If we choose
$\delta_A=2\Omega_A$ and $|\delta_A|\gg g_A$, then we can realize
the JC interaction in cavity A, which is described by Eq.
(\ref{JC}). However, in cavity B, we need the AJC interaction, which
requires the relation $\delta_B=-2\Omega_B$ and $|\delta_B|\gg g_B$
be chosen. Then we can bring the Hamiltonian (\ref{H_I}) to the AJC
interaction in the $\vert \pm\rangle$ atomic dressed basis
\begin{equation}
\label{AJC} H_B^{AJC}=\frac{g_B}{2}(\vert
-\rangle\langle+\vert\hat{a}_B+\vert
+\rangle\langle-\vert\hat{a}_B^\dag).
\end{equation}
The brief idea of producing the target entangled state is as
follows. We send a Rydberg atom preparing in state $\vert +\rangle$
into cavity A and make it undergo the dynamics governed by the JC
interaction. After an interaction time $t_A=\pi/(2g_A)$, the atom is
allowed to enter cavity B, undergoing the dynamics governed by the
AJC interaction of Eq. (\ref{AJC}). If we take $t_B=\pi/g_B$, the
final state for the cavity modes will be $1/\sqrt{2}(\vert
0\rangle_A\vert 0\rangle_B-\vert 1\rangle_A\vert 1\rangle_B)$ .

In order to illustrate the idea explicitly, we employ the time
evolution operator approach. Assume that at $t=0$, the system is in
state $\vert +\rangle\vert 0\rangle_A\vert 0\rangle_B$. At the stage
of the atom interacting with cavity A, the system is governed by the
interaction of Eq. (\ref{JC}). Then after an interaction time
$t_A=\pi/(2g_A)$ in cavity A, the atom-field state will be the same
as equation (\ref{entanglement2}). Subsequently, the atom enters
cavity B, and undergoes the dynamics of Eq. (\ref{AJC}). Then the
atom-field state will be
\begin{widetext}
\begin{eqnarray}
\frac{1}{\sqrt{2}}\vert-\rangle\vert0\rangle_A\vert0\rangle_B-\frac{i}{\sqrt{2}}(\cos(g_Bt/2)\vert
-\rangle\vert 0\rangle_B-i\sin(g_Bt/2)\vert +\rangle\vert
1\rangle_B)\vert 1\rangle_A.
\end{eqnarray}
\end{widetext}
If the interaction time $t_B=\pi/g_B$ is taken, the final state will
be
\begin{eqnarray}
\frac{1}{\sqrt{2}}(\vert 0\rangle_A\vert 0\rangle_B-\vert
1\rangle_A\vert 1\rangle_B)\vert +\rangle.
\end{eqnarray}
The atomic state has been factorized out and the modes of two
distant cavities end up in the following entangled state
\begin{equation}
\label{Bell2} \frac{1}{\sqrt{2}}(\vert 0\rangle_A\vert
0\rangle_B-\vert 1\rangle_A\vert 1\rangle_B).
\end{equation}
This state is also a maximally entangled state. Together with state
(\ref{Bell1}), they form two of the well-known Bell states. These
states have both fundamental applications in quantum mechanics and
practical applications in quantum information processing.

It is necessary to analyze the proposal requirements. To realize
this scheme, a two-level atom needs to sequentially interact with
two distant cavities. To generate the entangled coherent states of
two cavity modes, the initial states of the atom and two cavities
have only to be the ground states. After the atom leaves cavity B, a
tripartite entangled state involving one microscopic and two
mesoscopic systems has been prepared. One needs to measure the
atomic state in the bare basis to obtain the entangled coherent
states between two cavities. Therefore, to realize this protocol
requires that the atom and two cavity modes should not decay during
this process. One can control the interaction time of the atom with
the two cavities in experiments and implement this proposal in the
strong coupling regime to meet the requirements. In fact, from the
expressions for the parameters $\alpha$ and $\beta$, we know that
the total time for preparing the target entangled states is
determined by both the interaction time $t_A,t_B$ and the coupling
strengths $g_A,g_B$. In the case of generating Bell states, the
preparation time should be within the decay time of the atom and the
two cavities as well. However, in this case the interaction time
(the preparation time) is related to the coupling strengths. One
needs to suitably control the interaction time to implement this
protocol. Another requirement for producing Bell states is at the
beginning of the experiment, the atom needs to be prepared in the
coherent superposition state $\vert +\rangle=1/\sqrt{2}(\vert
g\rangle+\vert e\rangle)$, which can be produced by applying a
$\pi/2$ pulse to the atom before entering the cavity
\cite{RMP-73-565}.

We consider some experimental matters. For a potential experimental
system and set of parameters in microwave resonators
\cite{RMP-73-565}, the atomic configuration could be realized in
Rydberg atoms. We choose the single photon dipole coupling strength
as $g_A\sim g_B=g/(2\pi)\sim50$ kHz \cite{RMP-73-565}. Then for
generating entangled coherent states, the preparation time is about
$T\sim0.06$ ms with averaged photon number $|\alpha|=|\beta|=5$ in
each cavity. In the case of producing Bell states, the preparation
time is about $T\sim0.02$ ms. These results are in line with the
current experimental setups. Resonators stable over $100$ ms have
been reported recently \cite{APL}. The radiative life time for
Rydberg atoms is about $T_a\sim 30$ ms \cite{RMP-73-565}. Therefore,
the time needed to complete the procedure is much shorter than the
decay time of the atom and the cavities.

In conclusion, we have proposed a scheme for the generation of
two-mode entangled states between two separated cavities. It relies
on a two-level atom sequentially  coupling to two high-Q cavities
with a strong classical driving field. We demonstrate that by
suitably choosing the intensities and detunings of the fields and
coherent control of the dynamics, several different entangled states
such as entangled coherent states and Bell states can be produced
between the two cavity modes. With presently available experimental
setups in cavity QED, the realization of this proposal is feasible.

This work was supported by the National Natural Science Foundation
of China and National Key Basic Research Program. P.L. acknowledges
the quite useful discussions with Hongyan Li.


\begin{thebibliography}{24}
\expandafter\ifx\csname
natexlab\endcsname\relax\def\natexlab#1{#1}\fi
\expandafter\ifx\csname bibnamefont\endcsname\relax
  \def\bibnamefont#1{#1}\fi
\expandafter\ifx\csname bibfnamefont\endcsname\relax
  \def\bibfnamefont#1{#1}\fi
\expandafter\ifx\csname citenamefont\endcsname\relax
  \def\citenamefont#1{#1}\fi
\expandafter\ifx\csname url\endcsname\relax
  \def\url#1{\texttt{#1}}\fi
\expandafter\ifx\csname urlprefix\endcsname\relax\def\urlprefix{URL
}\fi \providecommand{\bibinfo}[2]{#2}
\providecommand{\eprint}[2][]{\url{#2}}

\bibitem[{\citenamefont{Bell}(1965)}]{Bell}
\bibinfo{author}{\bibfnamefont{J.~S.} \bibnamefont{Bell}},
  \bibinfo{journal}{Physics (Long Island City, NY)}
  \textbf{\bibinfo{volume}{1}}, \bibinfo{pages}{195} (\bibinfo{year}{1965}).

\bibitem[{\citenamefont{Nielsen and Chuang}(2000)}]{quantum_information}
\bibinfo{author}{\bibfnamefont{M.~A.} \bibnamefont{Nielsen}} \bibnamefont{and}
  \bibinfo{author}{\bibfnamefont{I.~L.} \bibnamefont{Chuang}},
  \emph{\bibinfo{title}{Quantum Computation and Quantum Information}}
  (\bibinfo{publisher}{Cambridge University Press, Cambridge, UK},
  \bibinfo{year}{2000}).

\bibitem[{\citenamefont{Shor}(1997)}]{Shor}
\bibinfo{author}{\bibfnamefont{P.~W.} \bibnamefont{Shor}},
  \bibinfo{journal}{SIAM J. Comput.} \textbf{\bibinfo{volume}{26}},
  \bibinfo{pages}{1484} (\bibinfo{year}{1997}).

\bibitem[{\citenamefont{K.Grover}(1997)}]{Prl-79-325}
\bibinfo{author}{\bibfnamefont{L.}~\bibnamefont{K.Grover}},
  \bibinfo{journal}{Phys.\ Rev. Lett.} \textbf{\bibinfo{volume}{79}},
  \bibinfo{pages}{325} (\bibinfo{year}{1997}).

\bibitem[{\citenamefont{Bennett et~al.}(1993)\citenamefont{Bennett, Brassard,
  Crepeau, Jozsa, Peres, and Wootters}}]{Prl-70-1895}
\bibinfo{author}{\bibfnamefont{C.~H.} \bibnamefont{Bennett}},
  \bibinfo{author}{\bibfnamefont{G.}~\bibnamefont{Brassard}},
  \bibinfo{author}{\bibfnamefont{C.}~\bibnamefont{Crepeau}},
  \bibinfo{author}{\bibfnamefont{R.}~\bibnamefont{Jozsa}},
  \bibinfo{author}{\bibfnamefont{A.}~\bibnamefont{Peres}}, \bibnamefont{and}
  \bibinfo{author}{\bibfnamefont{W.~K.} \bibnamefont{Wootters}},
  \bibinfo{journal}{Phys.\ Rev. Lett.} \textbf{\bibinfo{volume}{70}},
  \bibinfo{pages}{1895} (\bibinfo{year}{1993}).

\bibitem[{\citenamefont{Ekert}(1991)}]{Prl-67-661}
\bibinfo{author}{\bibfnamefont{A.~K.} \bibnamefont{Ekert}},
  \bibinfo{journal}{Phys.\ Rev. Lett.} \textbf{\bibinfo{volume}{67}},
  \bibinfo{pages}{661} (\bibinfo{year}{1991}).

\bibitem[{\citenamefont{Scully and Zubairy}(1997)}]{quantum_optics}
\bibinfo{author}{\bibfnamefont{M.~O.} \bibnamefont{Scully}} \bibnamefont{and}
  \bibinfo{author}{\bibfnamefont{M.~S.} \bibnamefont{Zubairy}},
  \emph{\bibinfo{title}{Quantum optics}} (\bibinfo{publisher}{Cambridge
  University Press, Cambridge, UK}, \bibinfo{year}{1997}).

\bibitem[{\citenamefont{Kimble}(1998)}]{Kimble}
\bibinfo{author}{\bibfnamefont{H.~J.} \bibnamefont{Kimble}},
  \bibinfo{journal}{Phys.\ Scr.} \textbf{\bibinfo{volume}{T76}},
  \bibinfo{pages}{127} (\bibinfo{year}{1998}).

\bibitem[{\citenamefont{Raimond et~al.}(2001)\citenamefont{Raimond, Brune, and
  Haroche}}]{RMP-73-565}
\bibinfo{author}{\bibfnamefont{J.~M.} \bibnamefont{Raimond}},
  \bibinfo{author}{\bibfnamefont{M.}~\bibnamefont{Brune}}, \bibnamefont{and}
  \bibinfo{author}{\bibfnamefont{S.}~\bibnamefont{Haroche}},
  \bibinfo{journal}{Rev.\ Mod.\ Phys.} \textbf{\bibinfo{volume}{73}},
  \bibinfo{pages}{565} (\bibinfo{year}{2001}).

\bibitem[{\citenamefont{Mabuchi and Doherty}(2002)}]{Sci298}
\bibinfo{author}{\bibfnamefont{H.}~\bibnamefont{Mabuchi}} \bibnamefont{and}
  \bibinfo{author}{\bibfnamefont{A.~C.} \bibnamefont{Doherty}},
  \bibinfo{journal}{Science} \textbf{\bibinfo{volume}{298}},
  \bibinfo{pages}{1372} (\bibinfo{year}{2002}).

\bibitem[{\citenamefont{Rauschenbeutel
  et~al.}(2001)\citenamefont{Rauschenbeutel, Bertet, Osnaghi, Nogues, Brune,
  Raimond, and Haroche}}]{pra-64-050301}
\bibinfo{author}{\bibfnamefont{A.}~\bibnamefont{Rauschenbeutel}},
  \bibinfo{author}{\bibfnamefont{P.}~\bibnamefont{Bertet}},
  \bibinfo{author}{\bibfnamefont{S.}~\bibnamefont{Osnaghi}},
  \bibinfo{author}{\bibfnamefont{G.}~\bibnamefont{Nogues}},
  \bibinfo{author}{\bibfnamefont{M.}~\bibnamefont{Brune}},
  \bibinfo{author}{\bibfnamefont{J.~M.} \bibnamefont{Raimond}},
  \bibnamefont{and} \bibinfo{author}{\bibfnamefont{S.}~\bibnamefont{Haroche}},
  \bibinfo{journal}{Phys.\ Rev. A} \textbf{\bibinfo{volume}{64}},
  \bibinfo{pages}{050301(R)} (\bibinfo{year}{2001}).

\bibitem[{\citenamefont{Wilk et~al.}(2007)\citenamefont{Wilk, Webster, Kuhn,
  and Rempe}}]{Sci317}
\bibinfo{author}{\bibfnamefont{T.}~\bibnamefont{Wilk}},
  \bibinfo{author}{\bibfnamefont{S.~C.} \bibnamefont{Webster}},
  \bibinfo{author}{\bibfnamefont{A.}~\bibnamefont{Kuhn}}, \bibnamefont{and}
  \bibinfo{author}{\bibfnamefont{G.}~\bibnamefont{Rempe}},
  \bibinfo{journal}{Science} \textbf{\bibinfo{volume}{317}},
  \bibinfo{pages}{488} (\bibinfo{year}{2007}).

\bibitem[{\citenamefont{Brune et~al.}(1996)\citenamefont{Brune, Hagley, Dreyer,
  Maitre, Maali, Wunderlich, Raimond, and Haroche}}]{Prl-77}
\bibinfo{author}{\bibfnamefont{M.}~\bibnamefont{Brune}},
  \bibinfo{author}{\bibfnamefont{E.}~\bibnamefont{Hagley}},
  \bibinfo{author}{\bibfnamefont{J.}~\bibnamefont{Dreyer}},
  \bibinfo{author}{\bibfnamefont{X.}~\bibnamefont{Maitre}},
  \bibinfo{author}{\bibfnamefont{A.}~\bibnamefont{Maali}},
  \bibinfo{author}{\bibfnamefont{C.}~\bibnamefont{Wunderlich}},
  \bibinfo{author}{\bibfnamefont{J.~M.} \bibnamefont{Raimond}},
  \bibnamefont{and} \bibinfo{author}{\bibfnamefont{S.}~\bibnamefont{Haroche}},
  \bibinfo{journal}{Phys.\ Rev. Lett.} \textbf{\bibinfo{volume}{77}},
  \bibinfo{pages}{4887} (\bibinfo{year}{1996}).

\bibitem[{\citenamefont{Monroe et~al.}(1996)\citenamefont{Monroe, Meekhof,
  King, and Wineland}}]{Sci272}
\bibinfo{author}{\bibfnamefont{C.}~\bibnamefont{Monroe}},
  \bibinfo{author}{\bibfnamefont{D.~M.} \bibnamefont{Meekhof}},
  \bibinfo{author}{\bibfnamefont{B.~E.} \bibnamefont{King}}, \bibnamefont{and}
  \bibinfo{author}{\bibfnamefont{D.~J.} \bibnamefont{Wineland}},
  \bibinfo{journal}{Science} \textbf{\bibinfo{volume}{272}},
  \bibinfo{pages}{1131} (\bibinfo{year}{1996}).

\bibitem[{\citenamefont{Agarwal and Puri}(1989)}]{pra-40}
\bibinfo{author}{\bibfnamefont{G.~S.} \bibnamefont{Agarwal}} \bibnamefont{and}
  \bibinfo{author}{\bibfnamefont{R.~R.} \bibnamefont{Puri}},
  \bibinfo{journal}{Phys.\ Rev. A} \textbf{\bibinfo{volume}{40}},
  \bibinfo{pages}{5179} (\bibinfo{year}{1989}).

\bibitem[{\citenamefont{Solano et~al.}(2003)\citenamefont{Solano, Agarwal, and
  Walther}}]{Prl-90-027903}
\bibinfo{author}{\bibfnamefont{E.}~\bibnamefont{Solano}},
  \bibinfo{author}{\bibfnamefont{G.~S.} \bibnamefont{Agarwal}},
  \bibnamefont{and} \bibinfo{author}{\bibfnamefont{H.}~\bibnamefont{Walther}},
  \bibinfo{journal}{Phys.\ Rev. Lett.} \textbf{\bibinfo{volume}{90}},
  \bibinfo{pages}{027903} (\bibinfo{year}{2003}).

\bibitem[{\citenamefont{Browne and Plenio}(2003)}]{pra-67-012325}
\bibinfo{author}{\bibfnamefont{D.~E.} \bibnamefont{Browne}} \bibnamefont{and}
  \bibinfo{author}{\bibfnamefont{M.~B.} \bibnamefont{Plenio}},
  \bibinfo{journal}{Phys.\ Rev. A} \textbf{\bibinfo{volume}{67}},
  \bibinfo{pages}{012325} (\bibinfo{year}{2003}).

\bibitem[{\citenamefont{Larson and Andersson}(2005)}]{pra-71-053814}
\bibinfo{author}{\bibfnamefont{J.}~\bibnamefont{Larson}} \bibnamefont{and}
  \bibinfo{author}{\bibfnamefont{E.}~\bibnamefont{Andersson}},
  \bibinfo{journal}{Phys.\ Rev. A} \textbf{\bibinfo{volume}{71}},
  \bibinfo{pages}{053814} (\bibinfo{year}{2005}).

\bibitem[{\citenamefont{Garcia-Maraver
  et~al.}(2006)\citenamefont{Garcia-Maraver, Eckert, Corbalan, and
  Mompart}}]{pra-74-031801}
\bibinfo{author}{\bibfnamefont{R.}~\bibnamefont{Garcia-Maraver}},
  \bibinfo{author}{\bibfnamefont{K.}~\bibnamefont{Eckert}},
  \bibinfo{author}{\bibfnamefont{R.}~\bibnamefont{Corbalan}}, \bibnamefont{and}
  \bibinfo{author}{\bibfnamefont{J.}~\bibnamefont{Mompart}},
  \bibinfo{journal}{Phys.\ Rev. A} \textbf{\bibinfo{volume}{74}},
  \bibinfo{pages}{031801(R)} (\bibinfo{year}{2006}).

\bibitem[{\citenamefont{Shu et~al.}(2006)\citenamefont{Shu, Zou, Xiao, and
  Guo}}]{pra-74-044301}
\bibinfo{author}{\bibfnamefont{J.}~\bibnamefont{Shu}},
  \bibinfo{author}{\bibfnamefont{X.~B.} \bibnamefont{Zou}},
  \bibinfo{author}{\bibfnamefont{Y.~F.} \bibnamefont{Xiao}}, \bibnamefont{and}
  \bibinfo{author}{\bibfnamefont{G.~C.} \bibnamefont{Guo}},
  \bibinfo{journal}{Phys.\ Rev. A} \textbf{\bibinfo{volume}{74}},
  \bibinfo{pages}{044301} (\bibinfo{year}{2006}).

\bibitem[{\citenamefont{Li et~al.}()\citenamefont{Li, Gu, Gong, and
  Guo}}]{eprint}
\bibinfo{author}{\bibfnamefont{P.~B.} \bibnamefont{Li}},
  \bibinfo{author}{\bibfnamefont{Y.}~\bibnamefont{Gu}},
  \bibinfo{author}{\bibfnamefont{Q.~H.} \bibnamefont{Gong}}, \bibnamefont{and}
  \bibinfo{author}{\bibfnamefont{G.~C.} \bibnamefont{Guo}},
  \eprint{arXiv:quant-ph/07111651}.

\bibitem[{\citenamefont{Jaynes and Cummings}(1963)}]{JC}
\bibinfo{author}{\bibfnamefont{E.~T.} \bibnamefont{Jaynes}} \bibnamefont{and}
  \bibinfo{author}{\bibfnamefont{F.~W.} \bibnamefont{Cummings}},
  \bibinfo{journal}{Proc. IEEE} \textbf{\bibinfo{volume}{51}},
  \bibinfo{pages}{89} (\bibinfo{year}{1963}).

\bibitem[{\citenamefont{Einstein et~al.}(1935)\citenamefont{Einstein, Podolsky,
  and Rosen}}]{epr}
\bibinfo{author}{\bibfnamefont{A.}~\bibnamefont{Einstein}},
  \bibinfo{author}{\bibfnamefont{B.}~\bibnamefont{Podolsky}}, \bibnamefont{and}
  \bibinfo{author}{\bibfnamefont{N.}~\bibnamefont{Rosen}},
  \bibinfo{journal}{Phys.\ Rev.} \textbf{\bibinfo{volume}{47}},
  \bibinfo{pages}{777} (\bibinfo{year}{1935}).

\bibitem[{\citenamefont{Kuhr et~al.}(2007)\citenamefont{Kuhr, Gleyzes, Guerlin,
  Bernu, Hoff, Del¨¦glise, Osnaghi, Brune, Raimond, Haroche et~al.}}]{APL}
\bibinfo{author}{\bibfnamefont{S.}~\bibnamefont{Kuhr}},
  \bibinfo{author}{\bibfnamefont{S.}~\bibnamefont{Gleyzes}},
  \bibinfo{author}{\bibfnamefont{C.}~\bibnamefont{Guerlin}},
  \bibinfo{author}{\bibfnamefont{J.}~\bibnamefont{Bernu}},
  \bibinfo{author}{\bibfnamefont{U.~B.} \bibnamefont{Hoff}},
  \bibinfo{author}{\bibfnamefont{S.}~\bibnamefont{Del¨¦glise}},
  \bibinfo{author}{\bibfnamefont{S.}~\bibnamefont{Osnaghi}},
  \bibinfo{author}{\bibfnamefont{M.}~\bibnamefont{Brune}},
  \bibinfo{author}{\bibfnamefont{J.-M.} \bibnamefont{Raimond}},
  \bibinfo{author}{\bibfnamefont{S.}~\bibnamefont{Haroche}},
  \bibnamefont{et~al.}, \bibinfo{journal}{Appl.\ Phys.\ Lett.}
  \textbf{\bibinfo{volume}{90}}, \bibinfo{pages}{164101}
  (\bibinfo{year}{2007}).

\end{thebibliography}

\end{document}